\documentclass[%
 reprint,
 amsmath,amssymb,
 aps,
]{revtex4-2}

\usepackage{graphicx}
\usepackage{dcolumn}
\usepackage{bm}

\begin{document}

\preprint{APS/123-QED}

\title{Minimal droplet shape representation in experimental microfluidics using Fourier series and autoencoders}

\author{Mihir Durve}
\email{mihir.durve@iit.it}
\affiliation{Center for Life Nano- \& Neuro-Science, Italian Institute of Technology (IIT), viale Regina Elena 295, Rome, 00161, Italy}

\author{Jean-Michel Tucny}
\affiliation{Center for Life Nano- \& Neuro-Science, Italian Institute of Technology (IIT), viale Regina Elena 295, Rome, 00161, Italy}
\affiliation{Dipartimento di Ingegneria, Università degli Studi Roma tre, via Vito Volterra 62, Rome, 00146, Italy}

\author{Sibilla Orsini}
\affiliation{NEST, Istituto Nanoscienze-CNR and Scuola Normale Superiore, Piazza San Silvestro 12, Pisa, 56127,  Italy}
\affiliation{Istituto per le Applicazioni del Calcolo del Consiglio Nazionale delle Ricerche, via dei Taurini 19, Roma, 00185, Italy}

\author {Adriano Tiribocchi}
\affiliation{Istituto per le Applicazioni del Calcolo del Consiglio Nazionale delle Ricerche, via dei Taurini 19, Roma, 00185, Italy}
\affiliation{INFN "Tor Vergata" Via della Ricerca Scientifica 1, 00133 Roma, Italy}

\author{Andrea Montessori}
\affiliation{Dipartimento di Ingegneria, Università degli Studi Roma tre, via Vito Volterra 62, Rome, 00146, Italy}

\author{Marco Lauricella}
\affiliation{Istituto per le Applicazioni del Calcolo del Consiglio Nazionale delle Ricerche, via dei Taurini 19, Roma, 00185, Italy}

\author{Andrea Camposeo}
\affiliation{NEST, Istituto Nanoscienze-CNR and Scuola Normale Superiore, Piazza San Silvestro 12, Pisa, 56127,  Italy}

\author{Dario Pisignano}
\affiliation{NEST, Istituto Nanoscienze-CNR and Scuola Normale Superiore, Piazza San Silvestro 12, Pisa, 56127,  Italy}
\affiliation{ Dipartimento di Fisica, Università di Pisa, Largo B. Pontecorvo 3, Pisa, 56127, Italy}

\author{Sauro Succi}
\affiliation{Center for Life Nano- \& Neuro-Science, Italian Institute of Technology (IIT), viale Regina Elena 295, Rome, 00161, Italy}
\affiliation{Department of Physics, Harvard University, 17 Oxford St, Cambridge, MA 02138, United States}

\begin{abstract}
We introduce a two-step, fully reversible process designed to project the outer shape of a generic droplet onto a lower-dimensional space. The initial step involves representing the droplet's shape as a Fourier series. Subsequently, the Fourier coefficients are reduced to lower-dimensional vectors by using autoencoder models.  The exploitation of the domain knowledge of the droplet shapes allows us to map generic droplet shapes to just 2D space in contrast to previous direct methods involving autoencoders that could map it on minimum 8D space. This 6D reduction in the dimensionality of the droplet's description opens new possibilities for applications, such as automated droplet generation via reinforcement learning, the analysis of droplet shape evolution dynamics and the prediction of droplet breakup. Our findings underscore the benefits of incorporating domain knowledge into autoencoder models, highlighting the potential for increased accuracy in various other scientific disciplines.
\end{abstract}

\keywords{Autoencoders, Microfluidics, Droplets, Minimal shape}

\maketitle

\section{Introduction}

The morphology of deformable structures, such as cells, small organisms, and droplets, often holds vital insights into their functionality \cite{cell_shape2,drop_shape,Denkov2015}. In the case of red blood cells, for example, deviations from their typical spherical shape result in impaired mobility and reduced oxygen-carrying capacity, leading to conditions like sickle cell anemia \cite{Lonergan}. Droplet shape also plays a key role in a variety of sectors of interest in microfluidics, such as drug delivery \cite{pais,pontrelli}, food and nutrition delivery \cite{Schroen2021} and tissue engineering \cite{guzowski,tiribocchi,montessori,sh_au}, to name a few ones.
However, the process of precise droplet generation with a desired shape for various applications is a challenging 
task due to time-dependent stages, from nucleation to detachment, and factors affecting the morphology, such as fluid properties, channel geometry, hydrodynamic instabilities and external perturbations. 

In practical applications, achieving high-throughput droplet generation with a predefined geometry is essential to meet industrial demands \cite{Kamperman2018,Jieke2023}. For instance, microdroplets with Janus-like properties (used, for example, in consumer electronics) or microcapsules for drug delivery often require production rates on the order of a few milliliters per minute, resulting in the generation of a few thousand droplets per minute. At such scales, the high throughput but precise droplet generation can be automated by machine learning based algorithms.

The objective of automated droplet production can be delineated into two primary sub-tasks, i) analysis of generated droplets and ii) feedback-based regulation of the droplet generation process. Recently, computer vision algorithms have been employed to address the first sub-task by analyzing droplet images. Specifically, algorithms based on YOLO (You Only Look Once) and DeepSORT (Deep Simple Online and Realtime Tracking) have been utilized to extract droplet trajectories and quantify parameters such as packing fraction, droplet ordering, and droplet count \cite{durve_2024}. This method can be further extended to achieve image segmentation, enabling the delineation of individual droplet boundaries from two-dimensional images captured during experimental or industrial operations.

For the second sub-task, reinforcement learning (RL) methods \cite{Sutton1998} have been proposed as effective tools for controlling the droplet production process \cite{Dressler}. The goal of reinforcement learning in this context is to enable the agent, i.e. the droplet generator, to learn an optimal policy ($\pi(a|s)$) that maximizes the cumulative reward, where the policy $\pi(a|s)$ determines the most suitable action $(a)$ to be taken in a given state $(s)$. Within the framework of droplet production, the action corresponds to adjustments made to the control parameters of the droplet generator, such as flow rates or chemical composition, which subsequently influence the morphology of the generated droplets, while the state $(s)$ is defined by the morphological characteristics of the droplets, for instance, the aspect ratio of a droplet in a 2D image 
as a simple state descriptor. The reward function is 
the deviation between the desired and the actual morphology of the produced droplets.

However, determining the droplet shapes typically observed in microfluidic  experiments requires more complex state descriptors than an aspect ratio, since the shapes significantly depart from spherical or ellipsoidal geometries.  
Since reinforcement learning algorithms, like Q-learning \cite{Watkins1992}, often encounter 
significant computational challenges for such a complex state space
represented by high-dimensional vectors, it is necessary, for practical implementation of these algorithms,
projecting the state space into a lower-dimensional representation.
In this respect, Khor et al \cite{khor} employed a convolutional autoencoder model to transform a given droplet shape into a low-dimensional vector representation, which was utilized to forecast whether the droplet would undergo instability and subsequent break-up. Their research indicated that an 8-dimensional shape descriptor provided the smallest representation capable of accurately reconstructing the original droplet shapes.

In this study, we present a novel two-step approach, recently demonstrated to map simple droplet shapes to 2-dimesional vectors, to further reduce the dimensionality of the vector representing droplet shapes within dense emulsions such as seen in Ref. \cite{Bogdan2022, MontessoriLangmuir}. Firstly, we characterize the droplet's contour using a Fourier series with a finite number of terms. Subsequently, we employ autoencoder neural networks to map the Fourier series coefficients onto a lower-dimensional vector space. Through the integration of Fourier series representation and autoencoder networks, we explore the tradeoff between the dimensionality of the mapping vector space and the accuracy of the regenerated shape. Our findings demonstrate that, with this fully reversible two-step procedure, a 2-dimensional vector space is adequate to achieve a mean square error of less than $10^{-4}$, indicating that the given droplet shape can be fully described by just two real numbers. Reducing the droplet shape to such low dimensional space paves a practical way to automate droplet production processes. 

The remainder of this article is structured as follows: Section \ref{section:expdrop} describes the droplets of which we are interested to recover the shape. Section \ref{section:TSEP} outlines the Fourier descriptors and the autoencoder neural network we use to compress the dimensionality of droplet shapes. We show the performance of our neural networks on four broad shape categories in Section \ref{section:Results}. We will then sum up our work in the Section \ref{section:Conclusion}.

\section{Microfluidic droplets}
\label{section:expdrop}

\subsection{Droplet generation procedure}
\label{dropgen}

\begin{figure}
    \centering
    \includegraphics[width=1\linewidth]{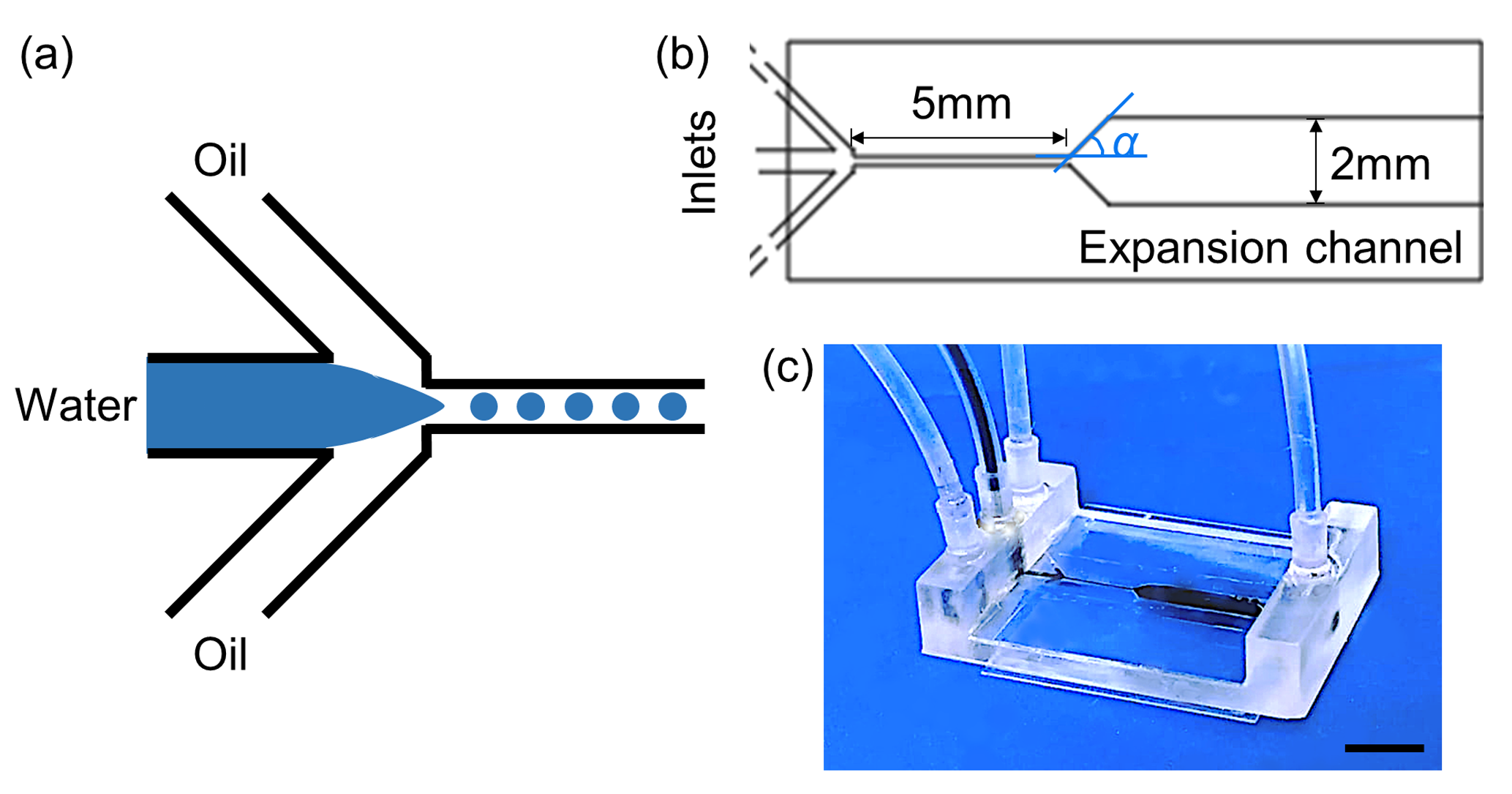}
    \caption{(a) Schematic illustration of the flow-focusing junction for W/O emulsion droplets generation. (b) Scheme of the device composed of the flow-focusing junction and the expansion channel, with aperture angle, $\alpha$.  (c) Photograph of a device. Scale bar: 5 mm.}
    \label{FigDevice}
\end{figure}

The droplet images analyzed in this study were produced using the experimental setup detailed in our previous work \cite{durve_2024}. Briefly, water-in-oil (W/O) emulsion droplets were created by using a flow-focusing microfluidic junction (Fig. \ref{FigDevice}a). A water solution with a black pigment was used as the dispersed phase and delivered to the device through a central inlet channel, while the continuous phase (sunflower oil) was injected through two lateral inlets. The generated W/O droplets were delivered to an expansion channel (width: 2 mm, length: 9 mm, depth: 500 $\mu$m, Fig. \ref{FigDevice}b,c), with opening angles $\alpha$ of either 30, 45, 60, or 90 degrees. The devices were fabricated by 3D printing, by using a digital light processing equipment (details about the fabrication of the devices are reported in Ref. \cite{durve_2024}). The arrays of droplets formed in the expansion channel were imaged by using a stereomicroscope and an high speed camera (Fastcam APX RS, Photron), at a rate of 3000 frames per second. By varying the opening angle $\alpha$ and adjusting the flow rates of the water and oil phases, we were able to produce droplets with a range of shapes and sizes (see Fig. \ref{Fig_drop_boundary}). 

\subsection{Shape extraction}
\label{subsection:shapex}

\begin{figure} [h]
\centering
\includegraphics[width=\linewidth]{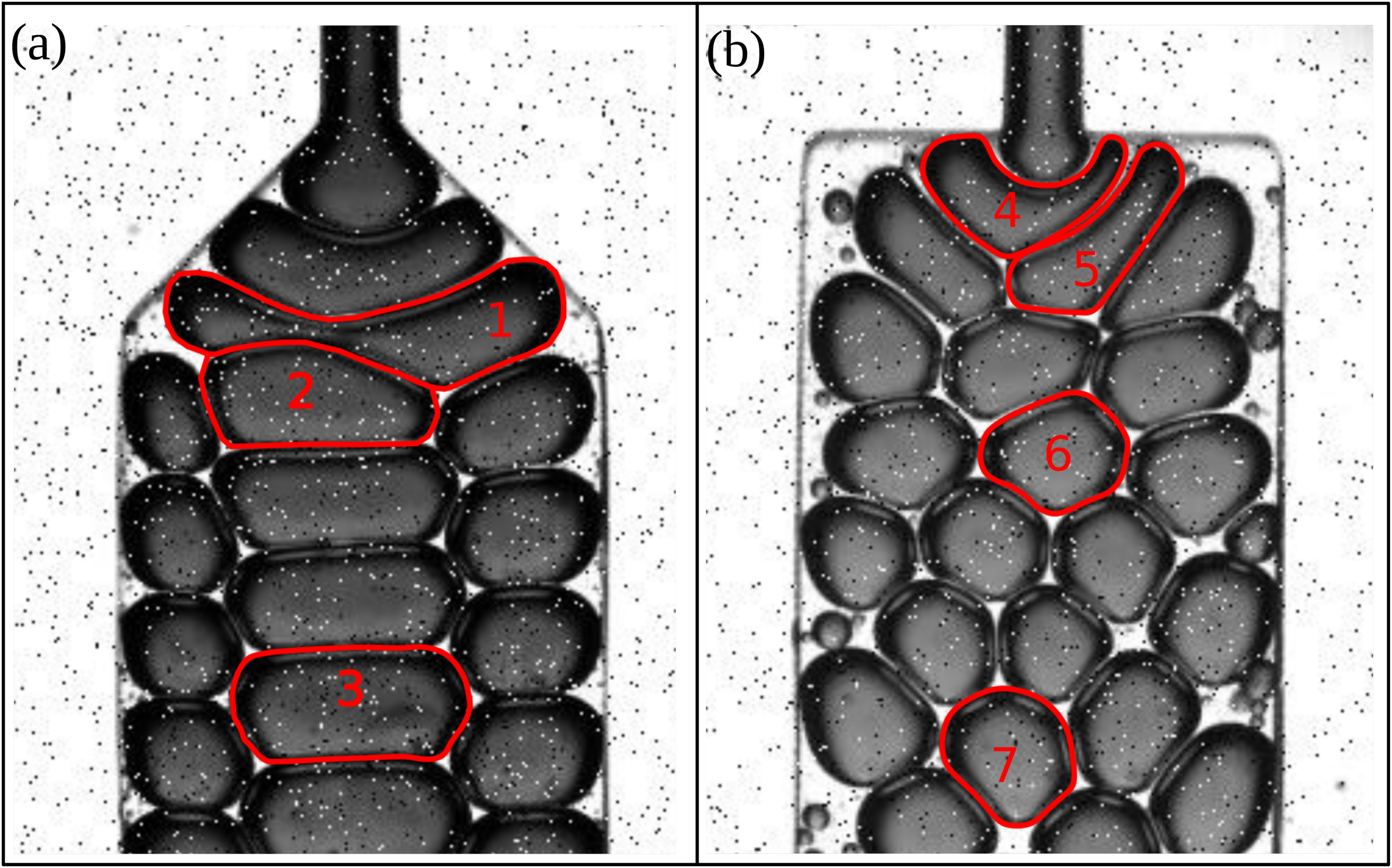}
\caption{ A few randomly selected droplet contours from the experimental realizations. The x-y coordinates of these contours are extracted manually and they are denoted as $f_o(t)$.  \label{Fig_drop_boundary}}
\end{figure}

To analyze the droplet shapes, we examined four videos of experimental realizations of dense emulsions, each characterized by different opening angles, $\alpha$, as detailed in Sec. \ref{dropgen}. We selected a subset of typical droplets, including those exhibiting significant deviations from the typical circular shape seen in free droplets (e.g., see droplet 1 in Fig. \ref{Fig_drop_boundary}). For these selected droplets, we manually delineated the boundary coordinates. Although, as mentioned before, advanced computer vision algorithms can be trained to automate the extraction of droplet boundaries, in this study we manually performed this task for ten droplets in the interest of time. Below we describe the two-step, fully reversible, process to map these droplet contours into low-dimensional vector space.

\section{Two-step encoding procedure}
\label{section:TSEP}

\begin{figure} [h]
\centering
\includegraphics[width=\linewidth]{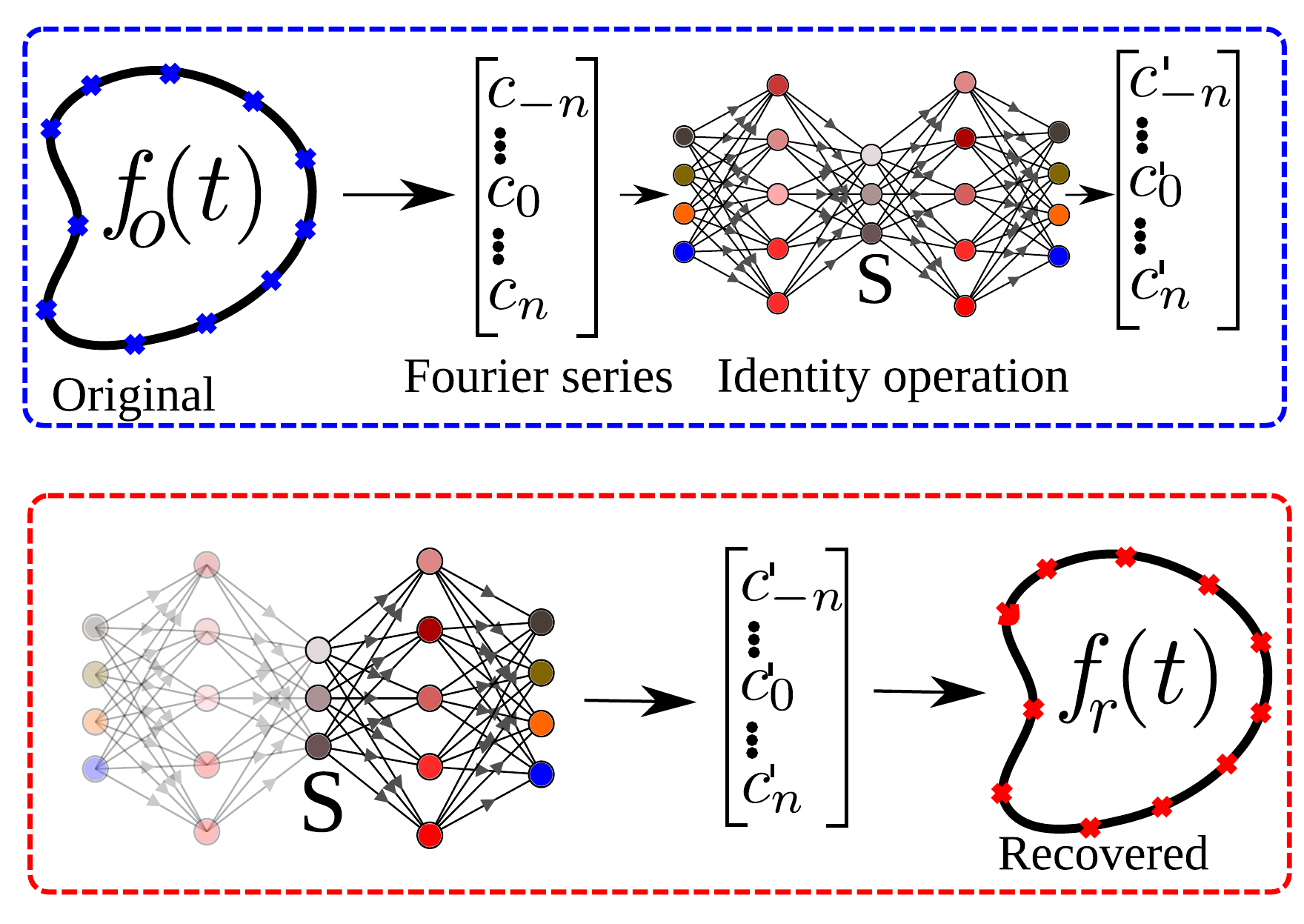}
\caption{Schematics of the method. During the first stage, the Fourier coefficients ($c_n$) of the droplet contour are mapped to lower dimensional vector $S$ with a neural network having bottleneck architecture. During the second phase, starting from $S$, the Fourier coefficients ($c_n'$) are recovered using the decoder part of the neural network and later used to recover the droplet contour $f_r(t)$. The goal of the training is to achieve $c_n'=c_n$. Figure adapted from Ref. \cite{durve2024dropletshaperepresentationusing}  \label{Fig_method}}
\end{figure}

The first step entails expressing a given droplet contour with the desired accuracy using a Fourier series with finite terms. Subsequently, in the second step, we employ an autoencoder model to compress the complex coefficients of the Fourier series into a low-dimensional vector space, denoted as $S$. A vector in $S$ encapsulates all the necessary information for reconstructing the contour. The procedural overview is illustrated in Figure \ref{Fig_method}. Below, we elaborate on the operational specifics of both steps.

\subsection{Fourier series description} 
\label{FSD}
With manual demarcation, we extract around 200 points for each droplet contour as shown in Fig. \ref{Fig_drop_boundary}.  The x,y coordinates of these points are approximated by the function $f_o(t)$ with the interval of parameter $t$ being [0,1]. In the complex form, the Fourier series representing $f_o(t)$ over this interval is given by:

\begin{equation}
 f_o(t) = \sum_{{n=-\infty}}^{n=+\infty} c_{n} e^{n \cdot 2\pi i t}.
 \label{eq_fourier}
\end{equation}

The coefficients {$c_n$} of this Fourier series are given by:

\begin{equation}
 c_n = \int_{0}^{1} f_o(t) e^{-n \cdot 2 \pi i t} dt,
 \label{eq_coefficient}
\end{equation}

where $c_0 = \int_{0}^{1} f_o(t) dt$ represents the average of $f_o(t)$ .

These coefficients $c_n$ are then computed numerically by integrating Eq. \ref{eq_coefficient} as follows:

\begin{equation}
 c_n = \int_{0}^{1} f_o(t) e^{-n \cdot 2 \pi i t} dt \approx \sum_{t=0}^{t=1} [f_o(t) e^{-n \cdot 2 \pi i t} \Delta t].
\end{equation}

Here, $\Delta t = 1/k$, where $k=200$ i.e. the number of extracted data points. For a given $n$, the coefficients $\{c_n\}$, are substituted back in Eq.\ref{eq_fourier} to reconstruct the droplet shape $f_r(t)$.

\begin{figure} [h]
\includegraphics[width=\linewidth]{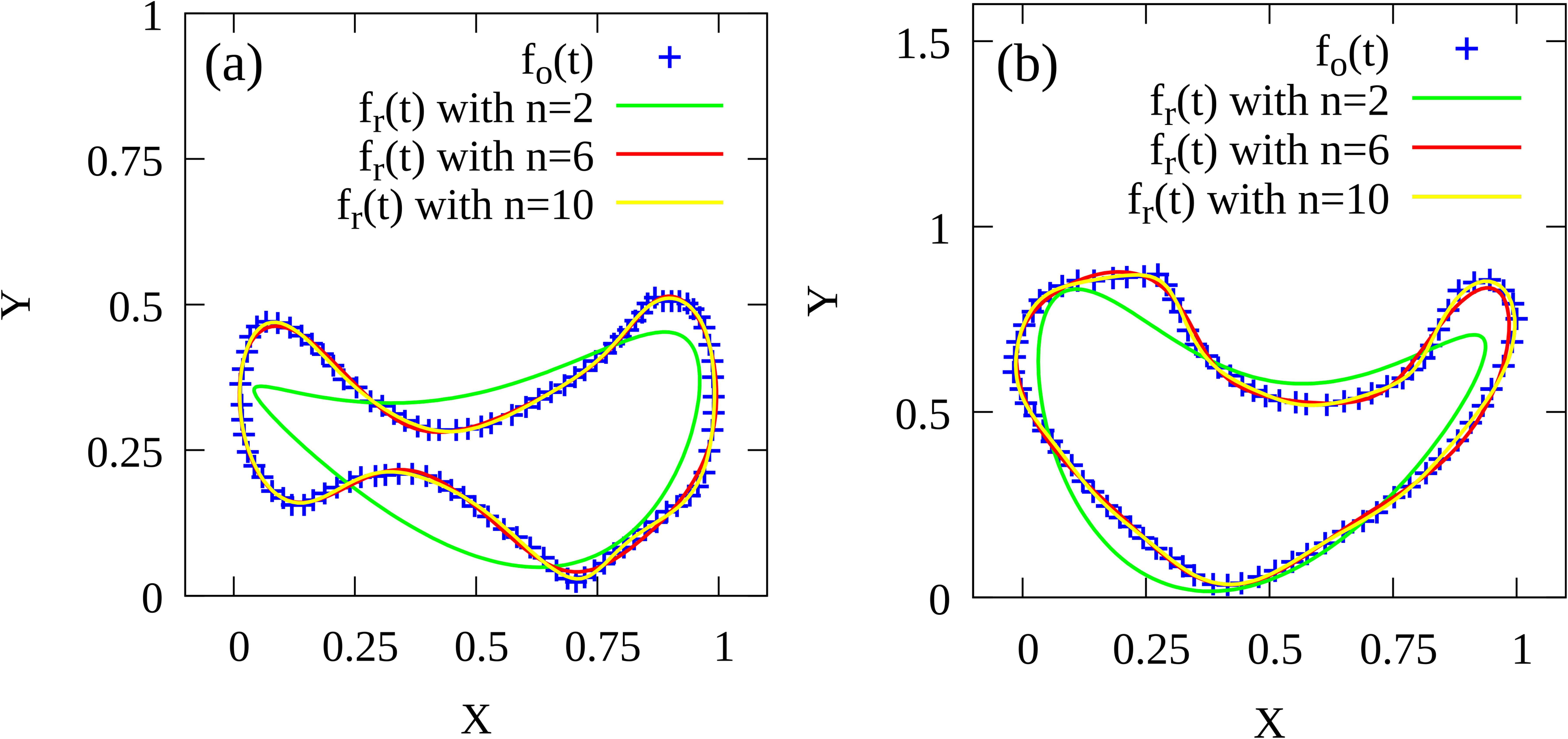}
\caption{ Two randomly selected droplet contours realized in microfluidic experiments. The droplets' contours ($f_o(t)$) marked manually are shown by blue crosses. (a) The contour of droplet 1 and (b) contour of droplet 4 from Fig. \ref{Fig_drop_boundary}. The contours ($f_r(t)$) generated by the Fourier series with an increasing number of modes $n$ in the series are shown by colored lines.}
\label{compare}
\end{figure}

In practice, we compute a finite number of Fourier modes $n$ that are enough to trace the contour of the droplet to a desired precision. As an illustration, we take contours $f_o(t)$ of droplets 1 and 4 from Fig. \ref{Fig_drop_boundary}. In Fig. \ref{compare}, we show the original contour $f_o(t)$ along with reconstructed contours $f_r(t)$ with an increasing number of Fourier modes $n$ in the series.  Here we observe that the mean square error (MSE) between $f_r(t)$ and $f_o(t)$ is less than $10^{-4}$ for $n \geq 6$, essentially making $f_r(t) \approx f_o(t)$ for practical purposes. In this work, we write the functional form of the droplet contours using $n=10$ Fourier modes. Thus each droplet is described by $2n+1=21$ complex coefficients or the corresponding 42 real numbers.   

\begin{figure} [h]
\centering
\includegraphics[width=1\linewidth]{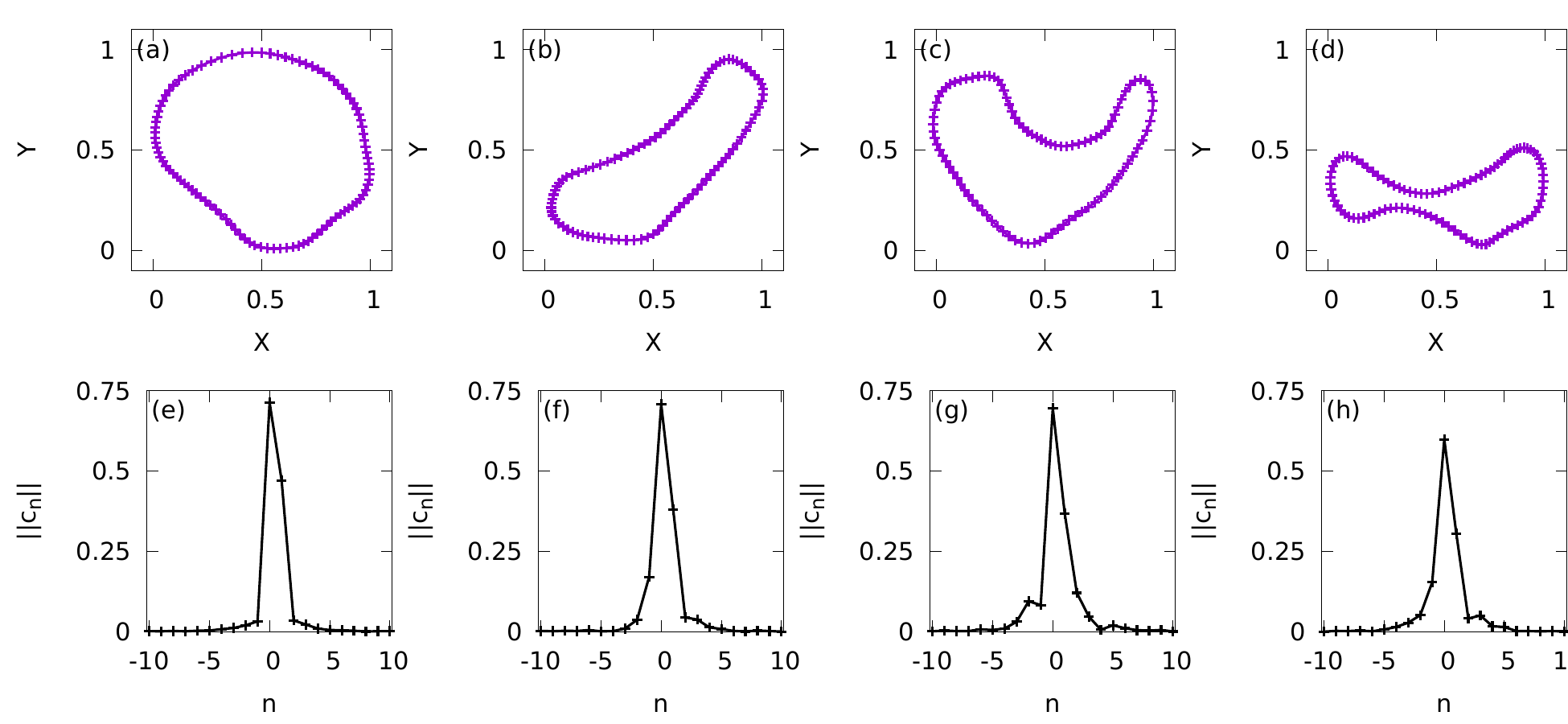}
\caption{(a)-(d) Randomly selected shapes taken from the experimental realization of droplets in microfluidic experiments. (e)-(h) Modulus of the coefficients ($||c_n||$) of the corresponding Fourier series denoting the contour. The total dataset consisted of such 20000 sets of Fourier series coefficients, each denoting a unique droplet. The total dataset is split in 9:1 proportion between the training and test datasets. \label{td}}
\end{figure}

In this study, we limit our attention to the ten droplet contours inferred through manual demarcation. The computed coefficients for these shapes, represented as $(c_n)$, provide a method for generating additional droplet contours. This is achieved by following a two-step procedure. In the first step, for each droplet, each coefficient $(c_n)$ undergoes multiplication by uniformly distributed numbers sampled from $U[1-\eta,1+\eta]$, where $\eta$ is numerically set to 0.1. Subsequently, in the second step, the resulting contour undergoes rotation by an angle $\theta$, randomly selected from the uniform distribution $U[-\pi,+\pi]$. 
Finally, the Fourier series with $n=10$ was utilized to generate training data for training the autoencoders. Thus, the training dataset was compiled by generating 20000 1-D vectors, each vector containing 42 real numbers corresponding to the ($2n+1$) complex coefficients of the Fourier series that in turn represent an individual droplet contour. A few such representative shapes and the modulus of their Fourier series coefficients are shown in Fig. \ref{td}.

In the next section, we describe the autoencoder network architecture and the training process using the generated training data as described above.

\subsection{Autoencoder network} 
\label{subsection:Nad}

As previously mentioned, each droplet contour is represented by a functional form defined by 42 real numbers. Autoencoder models are employed to compress these 42 numbers further, leveraging the correlation between the coefficients. We developed and trained eight autoencoder neural networks, all sharing an identical architectural framework except for the number of neurons in their bottleneck layers, which ranged from 1 to 8. Each network comprised 21 fully connected feedforward layers. The total count of trainable parameters varied accordingly: from 253,595 for the network with a single neuron in the bottleneck layer to 253,966 for the network with eight neurons in the bottleneck layer.
We designate the output at the bottleneck layer as the vector "S," which serves as a compressed representation of the Fourier coefficients provided at the input. Consequently, the dimensionality of vector $S$ (we call it the size of $S$) varies from 1 to 8 across these neural networks.

We partitioned the entire dataset, consisting of 20,000 one-dimensional (1D) vectors each with a length of 42, into training and test subsets in a 9:1 ratio. All the autoencoder networks were trained to perform identity operations, i.e. to produce the output that is identical to the input using in-house written python code. The Bayesian optimization method is used to find the best hyperparameters of the autoencoder network. The details of the hyper-parameter optimization and the detailed network architecture are described in the section below.

\section{Hyperparameter testing}


We employed the Keras Tuner library, specifically the \textit{BayesianOptimization} method \cite{omalley2019kerastuner}, to optimize the hyperparameters of our autoencoder-decoder network. BayesianOptimization uses Bayesian inference to predict the performance of different hyperparameter configurations and select the best one based on observed data. This method efficiently narrows down the search space to identify the optimal set of hyperparameters such as the optimizer, activation function and the number of neurons per layers. The objective was to minimize the mean square error of the validation loss, ensuring that the network not only compresses the information efficiently but also generalizes well. Early stopping was employed to prevent overfitting, with the patience parameter set to five epochs. 

The following activation functions were tested: ReLU (Rectified Linear Unit) \cite{Fukushima1969ReLU}, ELU (Exponential Linear Unit) \cite{clevert2016fast}, LeakyReLU (Leaky Rectified Linear Unit), PReLU (Parametric Rectified Linear Unit) \cite{xu2015empirical} and ThresholdedReLU \cite{IJAIN249}. These activation functions were applied consistently across all layers of the network. We also explored several optimizers to find the one that best suits our network’s training. The optimizers included in our search were Adam \cite{kingma2014adam}, SGD (Stochastic Gradient Descent) \cite{robbins1951stochastic}, RMSprop \cite{hinton2012rmsprop}, Nadam \cite{dozat2016incorporating}, and Lion \cite{chen2024symbolic}. Each optimizer has unique characteristics that influence convergence speed and stability, and we aimed to find the best optimizer for our Fourier coefficients compression task. The following bounds were set for the number of neurons per layer during the hyperparameter optimization. First layer: 50 to 250 neurons, in steps of 50; second layer: 100 to 250 neurons, in steps of 50; third and forth layer: 130 to 250 neurons, in steps of 40; fifth layer: 100 to 250 neurons, in steps of 50, and the sixth to ninth layers were kept with fixed number of neurons. These bounds were selected to avoid architectures we determined to be clearly suboptimal from initial tests. This search space allowed the optimization process to explore a wide range of network complexities, ultimately selecting the configuration that best captures the essential features of the input data. The best network configuration and the hyperparameters are mentioned in Table \ref{table1}

\begin{table}[h!]
\centering
\begin{tabular}{ |p{5cm}||p{3cm}|}
 \hline
 \multicolumn{2}{|c|}{Hyperparameter values for the autoencoder training} \\
 \hline
 \hline
 Activation function  & PReLU    \\
 \hline
 Optimizer  & Nadam    \\
 \hline
 Learning rate&  \parbox{3cm}{Epoch $< 15$ : $2 \times 10^-3$\\ Epoch 15-20 : $2 \times 10^-4$\\ Epoch 21-25 : $1 \times 10^-5$ \\ Epoch $>$ 25:$1 \times 10^-6$ \\ }   \\
\hline
 Network configuration & \parbox{3cm}{Input: 42\\ Layer 1: 50\\ Layer 2: 100\\ Layer 3: 250\\ Layer 4: 250\\ Layer 5: 100\\ Layer 6: 10\\ Layer 7: Size of S \\ Layer 8: Size of S\\ Layer 9: Size of S \\ symmetric \\ .\\ .\\ .\\Output: 42 \\} \\
 \hline
\end{tabular}
\caption{ \label{table1} Table with the autoencoder network configuration and hyperparameters.}
\end{table}

\section{Results}
\label{section:Results}
\subsection{Training and the Loss function}

\begin{figure}[h]
\centering
\includegraphics[width=1\linewidth]{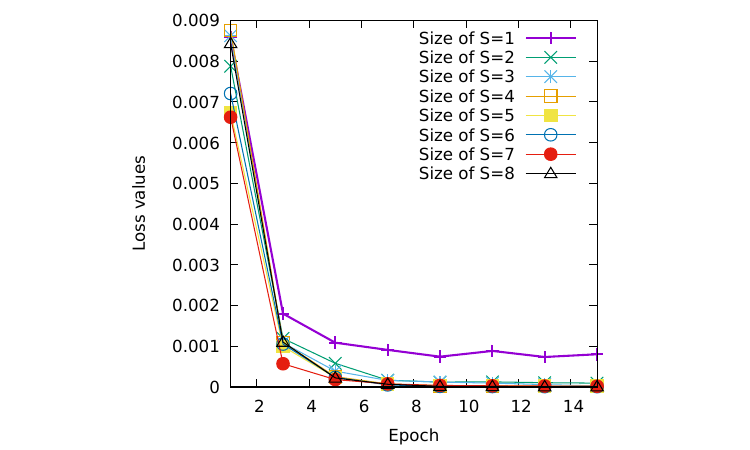}
\caption{ Loss function computed as the mean square error between the input and the output. Both the input and output are the $2\times(2N+1) = 42$ real numbers stemming from the coefficients of the Fourier series representing the droplet contour. Various curves represent neural networks with different numbers of neurons, which we call the "size of S", at the bottleneck layer.\label{loss_value}}
\end{figure}

All eight autoencoder models were trained using the dataset comprising 18000 instances and with a custom code developed in Python-TensorFlow. The code is made available on our institutional repository\cite{data_link}. The input consisted of Fourier series coefficients $c_n$. The autoencoder's task is to perform the identity operation. Hence the ideal output of the network is the same coefficients $c'_n = c_n$. However, practically, there is a small difference between the input numbers, i.e. $c_n$, and the output number $c'_n$. The details of the hyperparameters chosen for this training are mentioned in \ref{}. It is worthwhile to note that the training exercise is computationally inexpensive and is carried out on a laptop with an Intel-i7 CPU, 32 GB RAM, and no GPU. Each network takes about 20 mins to train and the inference takes about 10 milliseconds for each instance.

Fig. \ref{loss_value} shows the loss function as the training progress for all these eight autoencoders. We choose mean square error between the input and the output as the loss function in this work. As it can be seen, the loss values decrease and saturates at values $<0.0005$ for all the networks except the one with the "size of S=1". This network ensures slightly higher loss values, saturating at $\approx 0.001$ during the training.

\subsection{Error distribution}

\begin{figure} [h]
\centering
\includegraphics[width=1.0\linewidth]{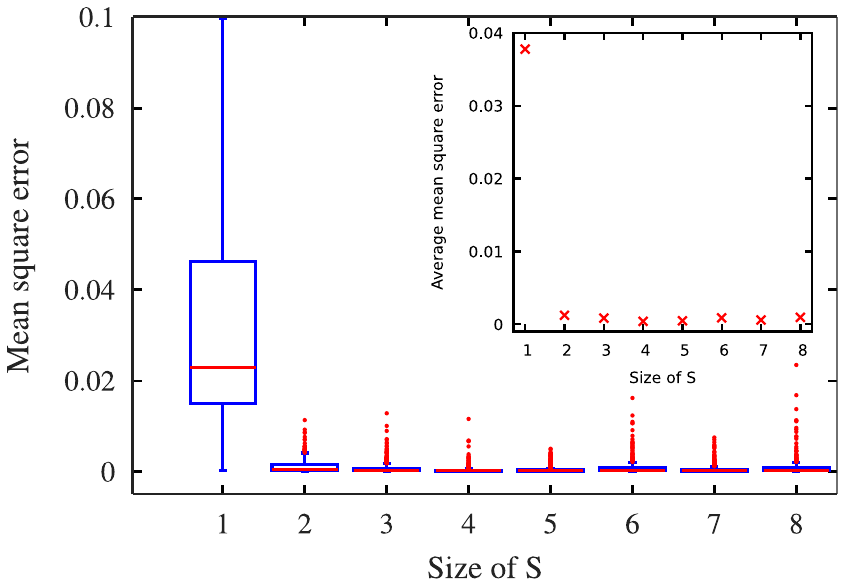}
\caption{Mean square error between the xy coordinates of the original curves $f_o(t)$ and the regenerated curves $f_r(t)$. The regenerated curves $f_r(t)$ are recovered starting from the vector $S$. The x-axis shows the neural network with the "Size of S" number of neurons at the bottleneck layer. Each boxplot shows statistics of mean square error computed over 1800 samples. In the insets, we show the average value of the mean square error between the original curves $f_o(t)$ and the regenerated curves $f_r(t)$. The mean square error is computed over 200 x-y coordinates for each droplet. \label{distribution}}
\end{figure}

Once the training was complete, test data was used to assess the quality of all the networks. Each network was given 2000 instances 
of Fourier series coefficients ($c_n$) as input and the corresponding values at the bottleneck layers (vector $S$) were saved for each input. These coefficients represent a droplet contour $f_o(t)$. Then, the values vector $S$ were given as input to the bottleneck layer to recover the Fourier coefficients ($c'_n$). The droplet contour was reconstructed using these $c'_n$ coefficients by the method described in Sec. \ref{FSD}. We call the reconstructed function $f_r(t)$. Practically, although the autoencoders are trained to take $c_n$ as input and produce $c'_n = c_n$ as output, there is a small difference between $c_n$ and $c'_n$ which results in a small difference between $f_o(t)$ and $f_r(t)$.

\begin{figure} [h]
\centering
\includegraphics[width=0.99\linewidth]{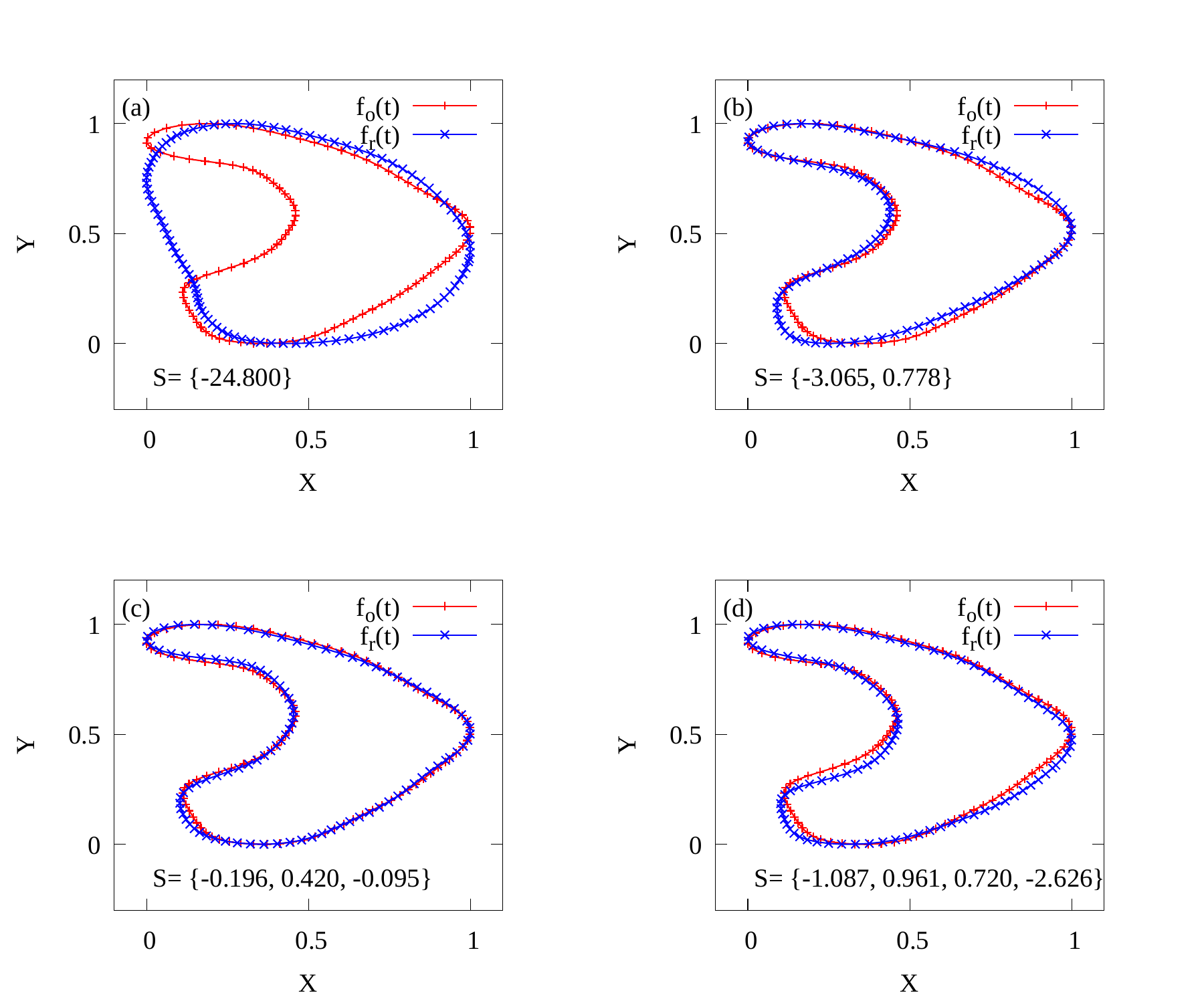}
\caption{Comparison between the original curve $f_o(t)$ in blue color and the recovered droplet contour $f_r(t)$ in red color starting from Vector $S$. The subplots show the comparison for increasing number of neurons at the bottleneck layer i.e. size of S = (a)1, (b)2, (c)3, (d)4. The minimal representation, i.e. the vector $S$, is shown at the bottom of each subplot truncated at the third decimal value. \label{compare2}}
\end{figure}

Figure \ref{distribution} illustrates the distribution of the mean square error (MSE) between the original contour $f_o(t)$ and regenerated contour $f_r(t)$ for 2000 instances in the test data for various networks. The mean square error is computed as
\begin{equation}
    MSE = \frac{1}{k} \sum_{i=1}^{k} ( (x_o(i)-x_r(i))^2 + (y_o(i)-y_r(i))^2,
\end{equation}
where, $x_o$ and $y_o$ are the points in $f_o(t)$, $x_r$ and $y_r$ are the points in $f_r(t)$ and each droplet contour is represented by $k=200$ coordinates. In the inset, we show the average of these mean square error (MSE) values over 2000 instances for each autoencoder network. The autoencoder with size of S $\geq 2$ performs equally well while the network with the size of S $=1$ endures large errors. In the next section, we show visual representation of a single instance of $f_o(t)$ and $f_r(t)$ to provide the qualitative sense of the mean square error numbers reported here.

\subsection{Comparison of original and regenerated droplets} 
 
An arbitrary instance from the test data is selected to assess the compression quality of the autoencoders. The same Fourier series coefficients were fed through the first four autoencoder networks, and the droplet contours $f_r(t)$ (red curve) recovered from the minimal representation $S$  are depicted alongside the original function $f_o(t)$ (blue curve) in Fig. \ref{compare2} for these four networks. In Fig. \ref{compare2} (b), (c), and (d), a nearly perfect overlap can be observed between the original contour and the regenerated contour. Given that the average mean error values are comparable for all networks where size of S $\geq 2$, we conclude that droplet contours can effectively be represented by only 2 real numbers. Starting from the minimal representation  $S$, the droplet shape recovered through the reversible process outlined in this study closely matches the original contour.   

\section{Conclusion}
\label{section:Conclusion}
Droplet shapes carry significant information about the external and internal forces acting upon them. Due to the inherent physical properties of liquids, these shapes form contours that are continuous and free from discontinuities and kinks. However, mapping generic droplet contours onto a minimal representation space is a complex task. Achieving a minimal representation of droplet shapes is highly desirable for various applications. For example, such minimal representations could facilitate the use of reinforcement learning algorithms to automate droplet generation processes. Additionally, representing droplet shapes as vectors could aid in predicting phenomena such as droplet coalescence or breakup within microfluidic channels. Previous research has demonstrated that direct input of droplet contours into autoencoder networks can compress the shape to an eight-dimensional vector.

In this study, we introduce a two-step process that leverages domain-specific knowledge about the smoothness of droplet contours, expressed through Fourier series, to map generic droplet contours effectively onto a two-dimensional space. Given that the task involves compressing the Fourier coefficients, the autoencoder employed in this process benefits from a significantly reduced number of nodes. This efficiency arises because the network can easily learn and exploit the inherent correlations between the Fourier coefficients. Essentially, this means that just two real numbers suffice to encapsulate the necessary information for reconstructing the droplet contour.

This minimal contour representation marks a significant advancement in various applications. For instance, reinforcement learning algorithms, such as Q-learning, become computationally feasible with a two-dimensional state space. Furthermore, the evolution of droplet shapes can be studied and easily visualized when mapped to a 2D space. 

However, it is worthwhile to note that this study has limitations regarding finding the connection between the minimal representation and the physical attributes of the droplets. Future research will aim to explore such relationships - if they exist.

\section{Acknowledgment}
The authors acknowledge funding from the European Research Council ERC-PoC2 grant No. 101081171 (DropTrack). J.-M. T. thanks the FRQNT “Fonds de recherche du Québec – Nature et technologies (FRQNT)" for financial support (Research Scholarship No. 314328).  M.L. acknowledges the support of the Italian National Group for Mathematical Physics (GNFM-INdAM)

\bibliography{version1}

\end{document}